\documentclass{article}
\usepackage{spconf}
\usepackage{multirow,amsmath,graphicx}
\usepackage{caption}
\usepackage{hyperref}
\usepackage{amsfonts}

\title{A context based deep learning approach for unbalanced medical image segmentation}

\name{
\begin{tabular}{@{}c@{}}
Balamurali Murugesan$^{1,2,\star}$\thanks{$^{\star}$ Contributed equally \newline $^{\dagger}$ Work done while interning at HTIC} \qquad Kaushik Sarveswaran$^{2,\star,\dagger}$ \qquad Vijaya Raghavan S $^{2,\dagger}$ \\ Sharath M Shankaranarayana$^{3}$ \qquad Keerthi Ram$^{2}$  \qquad Mohanasankar Sivaprakasam$^{1,2}$
\end{tabular}}

\address{$^{1}$ Indian Institute of Technology Madras (IITM), India \\ $^{2}$ Healthcare Technology Innovation Centre (HTIC), IITM,  India \\ $^{3}$ Zasti, India}

\begin{document}
%
\maketitle
\begin{abstract}
Automated medical image segmentation is an important step in many medical procedures. Recently, deep learning networks have been widely used for various medical image segmentation tasks, with U-Net and generative adversarial nets (GANs) being some of the commonly used ones. Foreground-background class imbalance is a common occurrence in medical images, and U-Net has difficulty in handling class imbalance because of its cross entropy (CE) objective function. Similarly, GAN also suffers from class imbalance because the discriminator looks at the entire image to classify it as real or fake. Since the discriminator is essentially a deep learning classifier, it is incapable of correctly identifying minor changes in small structures. To address these issues, we propose a novel context based CE loss function for U-Net, and a novel architecture Seg-GLGAN. The context based CE is a linear combination of CE obtained over the entire image and its region of interest (ROI). In Seg-GLGAN, we introduce a novel context discriminator to which the entire image and its ROI are fed as input, thus enforcing local context. We conduct extensive experiments using two challenging unbalanced datasets: PROMISE12 and ACDC. We observe that segmentation results obtained from our methods give better segmentation metrics as compared to various baseline methods. 
\end{abstract}
\section{Introduction}  
Automated medical image segmentation is a preliminary step in many medical procedures. Recently, deep learning methods were introduced into the context of medical image segmentation and have gained great popularity \cite{survey}. U-Net \cite{unet} is the most commonly used deep learning network for medical image segmentation. Even though U-Net is widely used, it has issues in dealing with labels having class imbalance. The issue is mainly due to the usage of cross entropy (CE) loss, as it is well known in literature that CE has difficulty in handling class imbalance. Since most of the medical images have fewer foreground pixels relative to larger background pixels, using CE will learn a decision boundary biased towards the majority class, which would result in inaccurate segmentation. 
\begin{figure}
    \centering
    \includegraphics[width=0.8\linewidth]{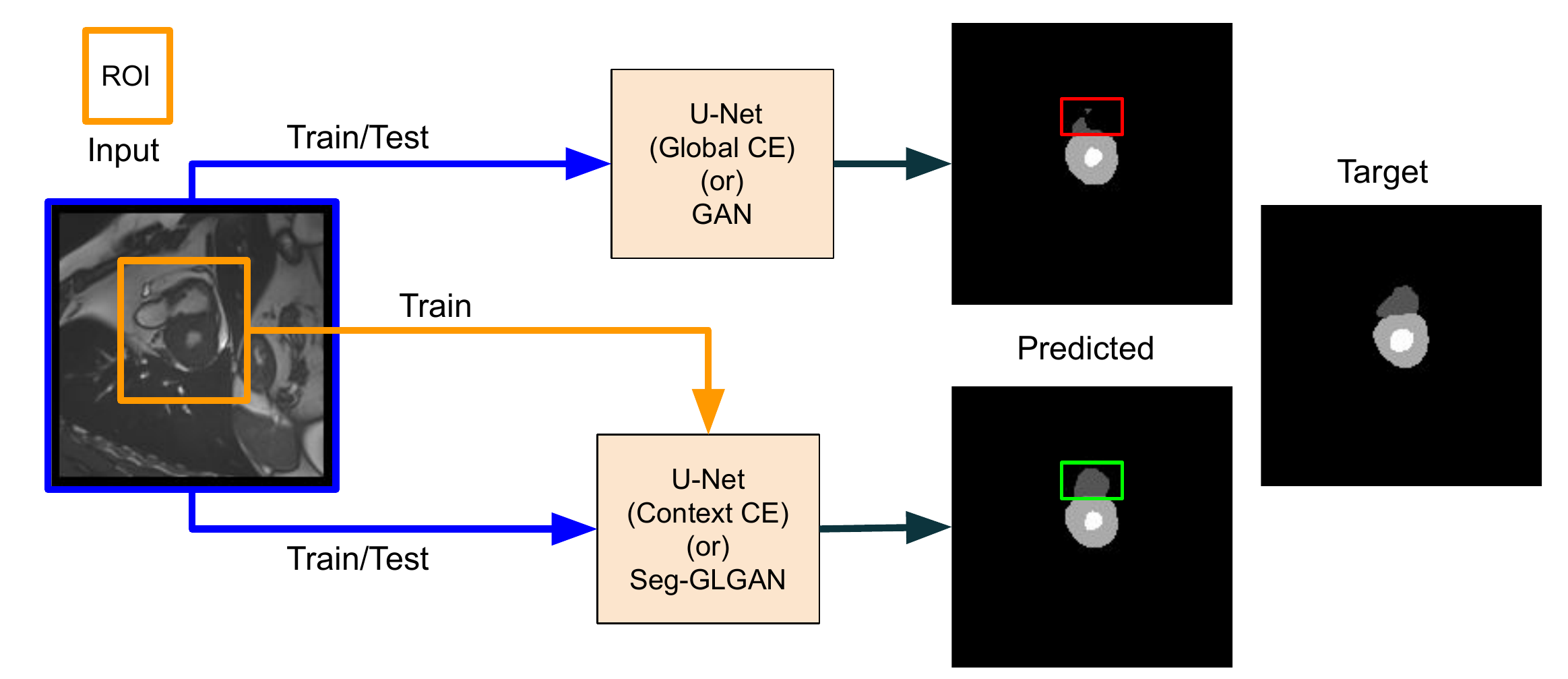}
    \caption{Illustration depicting the differences between the baseline and our proposed frameworks. During Training, while U-Net with Global Cross Entropy(CE) Loss/GAN networks take in as input the whole image only, our proposed frameworks(U-Net with Context-based CE Loss and Seg-GLGAN) also take in the cropped ROI as an additional input. Both the baseline and our proposed frameworks only take in the whole image during inference.}
    \label{fig:overview}
\end{figure}
In order to alleviate the class imbalance problem, the basic approach is to assign weights to classes based on the inverse of it occurrences \cite{unet}. This approach has two drawbacks: 1) assigning a proper weight will be an issue for a dataset with varying object sizes, 2) the least frequent class will be affected by noise and it may result in unstable training. The most commonly used loss functions to tackle class imbalance problem are dice coefficient \cite{dice} and focal loss \cite{focal_loss}. In focal loss, a modulating factor is added to the CE loss to differentiate between easy (background) and hard (foreground) examples. Both these loss functions have shown promising results compared to CE but faces difficulty with datasets having severe class imbalance.

Lately, Generative Adversarial Networks (GANs) \cite{nips_gan} have been extensively used for various challenging medical segmentation tasks. GAN consists of generator and discriminator; in the case of segmentation, the generator is the image to mask prediction network while discriminator is the real/fake classification network. In \cite{isbi_gan}, GANs are used for skin lesion segmentation task, and it has been shown that by adversarial training, the discriminator helps the generator to provide refined lesion segmentation masks. Similarly, GANs have also shown improved results in brain tumor segmentation \cite{brats_gan}. The joint segmentation of optic cup and disc have also been successfully studied using GANs \cite{cgan_sharath}. In the above discussed methods, the discriminator looks at the entire predicted/target label to classify it as real/fake. Because of this, the discriminator will be insensitive to minor changes in the local region as it lacks specific local information. Similarly, the CE loss function in the U-Net considers the entire predicted/target label without having any knowledge on the specific region of interest (ROI). Due to the lack of local information, the shape of predicted segmentation label by U-Net will be dissimilar to target label. Considering the drawbacks stated above, we have designed a loss function for U-Net and a novel architecture based on GAN which directly alleviates the class imbalance problem and produces relatively similar shape to ground truth by considering both global and local information. The overall approach is depicted in Fig. \ref{fig:overview}. The key contributions of our paper are summarized as follows: 


\begin{itemize}
    \item We propose a joint loss function called context based cross entropy (CE) loss for U-Net which is the linear combination of global and local CE loss. The global CE loss takes the entire predicted and target label, while the local CE loss takes the ROI of predicted and target label. 
    \item We propose a novel architecture Seg-GLGAN, which consists of a generator and a context discriminator. The context discriminator consists of global and local feature extractor. The global feature extractor takes the entire predicted/target label, while the local feature extractor takes the ROI of predicted/target label.  
    \item We also show that the local information can be passed to context based CE loss function in U-Net and context discriminator in Seg-GLGAN through two ways: 1) \textit{Static ROI}: the ROI is fixed to certain dimension and is used for the entire train dataset. 2) \textit{Dynamic ROI}: the ROI is varied for each image/label pair in train dataset based on the size of the object in the target label. 
    \item We conduct extensive experiments on two challenging and unbalanced segmentation datasets: 1) \textit{Binary segmentation}: Prostate MR Image Segmentation 2012 (PROMISE12) and 2) \textit{Multi-class segmentation}: Automated Cardiac Diagnosis Challenge (ACDC). We show that adding local information to U-Net and GAN gives substantial improvement over various baseline segmentation metrics.
\end{itemize}
\section{Methodology}
\subsection{U-Net: Proposed Context based Cross Entropy Loss}
U-Net uses the global CE loss ($\mathcal{L}_{global}$). The global CE loss is obtained between the entire target and predicted label. Because of the class imbalance, the minimization of the global CE loss doesn't imply that we have obtained the global minimum. So, we propose a local CE loss ($\mathcal{L}_{local}$) which is obtained between the ROI of the target and predicted label. The addition of $\mathcal{L}_{local}$ to $\mathcal{L}_{global}$ will result in better optimization. The minimization of $\mathcal{L}_{local}$ can only be achieved by doing a better segmentation in the ROI. This results in better overall segmentation. We call the proposed joint loss function as context based CE loss $\mathcal{L}_{context}$ and is given as: 

\begin{equation}
 \begin{aligned}
 \mathcal{L}_{context} &=  \mathcal{L}_{global} + \lambda \mathcal{L}_{local} \\
 \mathcal{L}_{global/local} &= - \sum_{\boldsymbol{i} \, \epsilon\, \Omega_{EI/ROI}}\sum_{c=1}^{C}y_{\boldsymbol{i}c} \, ln\, \hat{y}_{\boldsymbol{i}c}
\end{aligned}
\end{equation}

where $C$ denotes total number of classes, $\hat{y}_{\boldsymbol{i}c}$ denotes the predicted probability for true label $y_{\boldsymbol{i}c}$ after softmax activation function. $\mathcal{L}_{global}$ and $\mathcal{L}_{local}$ denotes the CE obtained over entire label ($\Omega_{EI}$) and ROI ($\Omega_{ROI}$) of the label. 

\subsection{Proposed Segmentation Global Local Generative Adversarial Network (Seg-GLGAN)}
\subsubsection{Generative Adversarial Networks}
GAN is used to learn a mapping from observed image $x$ to $y$, $G: x \rightarrow y$. In the case of segmentation, the generator G is trained jointly with discriminator D to produce segmentation label similar to target label through the following adversarial optimization function:
\begin{equation}
 L_{GAN}(G,D) = \mathbb{E}_{y}[\log D(y)] + \mathbb{E}_{x} [\log(1 - D(G(x))]   
\end{equation}
In general, the discriminator is a classifier network which extracts features from the entire input image and classifies it as either real or fake. For unbalanced class segmentation datasets, this approach will not be sufficient for the discriminator to distinguish the image as real or fake, resulting in coarse segmentation. An alternative approach in literature is the use of PatchGAN\cite{pix2pix} discriminator, which extracts features from the local patches and classifies the entire image as real or fake by equally averaging the patch responses; this does not explicitly take the context information into account, since every local region is weighted equally. We propose Seg-GLGAN network which considers both global and local context to provide a better understanding of real and fake images. 


\begin{figure*}
    \centering
    \includegraphics[width=0.8\linewidth]{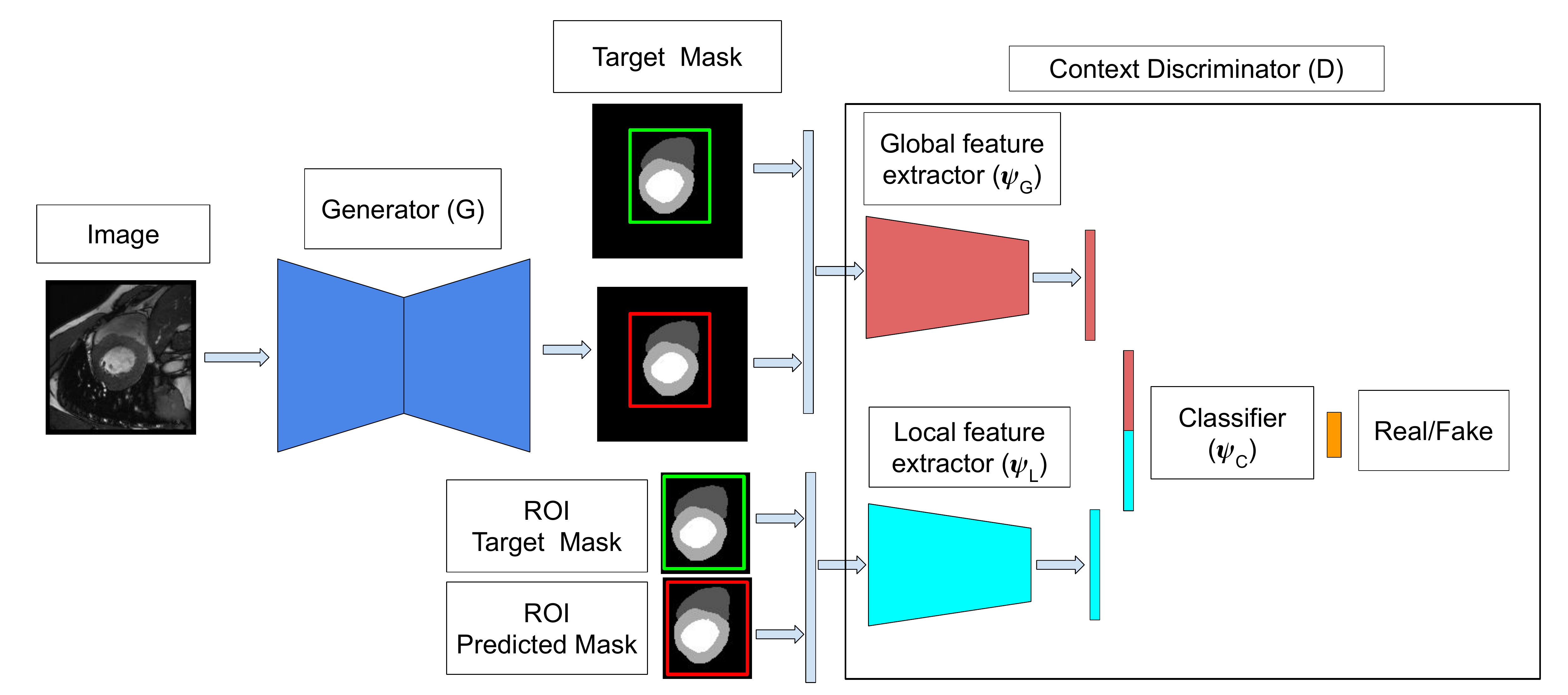}
    \caption{Outline of the proposed Segmentation Global Local Generative Adversarial Network (Seg-GLGAN)}
    \label{fig:archgan}
\end{figure*}

\subsubsection{Network architecture}
The proposed architecture Seg-GLGAN is inspired from a GAN based work on inpainting \cite{gl-gan}. In inpainting, the global and local information is used jointly to provide better context enabling a consistent restoration. The architecture Seg-GLGAN has a generator and context discriminator. The outline of the proposed architecture Seg-GLGAN is depicted in Fig. \ref{fig:archgan}. The generator is the most commonly used U-Net. The context discriminator consists of three components:

\textbf{Global feature extractor} ($\Psi_{G}$): The global feature extractor takes the whole probability or target segmentation mask as input and outputs a 64 dimensional feature vector. The global feature extractor has 3 convolutional layers followed by 2 fully connected layers. The activation function Leaky ReLu is used for each layer. Each convolution layer is followed by an average pooling layer. The kernel sizes of 9$\times$9, 5$\times$5 and 5$\times$5 are used for each convolution with output channels of 32, 64 and 64.

\textbf{Local feature extractor} ($\Psi_{L}$): The local feature extractor takes ROI of the probability or target segmentation mask and outputs a  64-dimensional feature vector. The architecture is largely similar to that of the global feature extractor except for the dimensions of the feature vector of the fully connected layer, which is modified according to the ROI dimensions.

\textbf{Classifier} ($\Psi_{C}$):
A 128-dimensional vector is formed by concatenating the feature vectors from $\Psi_{G}$ and $\Psi_{L}$. This vector is given as input to a single fully-connected layer followed by sigmoid, to output a confidence score for classifying the probability or target segmentation mask as real or fake. 

\vspace{-5mm}
\subsubsection{Loss function}
Let $\Phi$ be a function to extract ROI given an image. The joint optimization of generator and context discriminator for the proposed architecture Seg-GLGAN is given by:
\begin{equation}
    L_{Seg-GLGAN} = L_{MCE}(G) + \lambda L_{GAN}(G,D) 
\end{equation}
\begin{equation}
 \begin{aligned}
 L_{MCE}(G)  &= - \mathbb{E}_{x,y}\Big{[}\sum_{c=1}^{C}y_{c} \, ln\,G_{c}(x)\Big{]} \\ 
 L_{GAN}(G,D) &= \mathbb{E}_{y}[\ln D(y)] + \mathbb{E}_{x} [\ln(1 - D(G(x))]   \\ 
 D(y) &= \Psi_{C}(\Psi_{G}(y) || \Psi_{L}(\Phi(y))) \\
 D(G(x)) &= \Psi_{C}(\Psi_{G}(G(x)) || \Psi_{L}(\Phi(G(x))))
 \end{aligned}
\end{equation}

\section{Experiments and Results}
\vspace{-2mm}
\subsection{Dataset and Evaluation metrics}
We evaluate our proposed context based cross entropy loss for U-Net and the Seg-GLGAN architecture with two challenging datasets having unbalanced classes. 

\textbf{Prostate MR Image Segmentation 2012 (PROMISE12)}:
The PROMISE12 \cite{promise_dataset} dataset consists of 50 T2-weighted MR volumes of prostate and its corresponding binary segmentation mask with different resolutions. The volumes are resampled to an isotropic $1mm^{3}$, normalized to (0-1) and center cropped to $128\times128$. The volumes are randomly split into 70\% for training and 30\% for testing. 2D slices are extracted from each volume resulting in 1127 slices for training and 530 slices for validation. 

\textbf{Automated Cardiac Diagnosis Challenge (ACDC)}:
The ACDC \cite{acdc_dataset} dataset consists of 150 patient records of cardiac MR volume and its respective multi-class segmentation mask. The segmentation mask corresponds to Left Ventricle (LV), Right Ventricle (RV), Myocardium (Myo) and background. The volumes are normalized to (0-1). The patient records are randomly split into 70\% for training and 30\% for testing.  From the patient records, 2D slices are extracted and cropped to 160$\times$160. The extracted 2D slices amount to 1309 for training and 532 for validation.

\textbf{Evaluation metrics}:
Dice Similarity Coefficient (DICE) and Hausdorff Distance (HD) are used to evaluate the segmentation performance.
\subsection{Implementation Details}

\textbf{ROI selection}:
i) \textit{Static ROI}(StROI): During training, the dimensions of ROI for PROMISE12 and ACDC are empirically set to 50$\times$50 and 60$\times$60. ii) \textit{Dynamic ROI}(DyROI): The dimensions of the ROI is adapted dynamically based on the size of the object. In the case of Seg-GLGAN, the fully connected network in the context discriminator is replaced with Global Average Pooling to handle ROI with different dimensions. 

\textbf{Training details}
Models are implemented in PyTorch and code is publicly available \footnote{\href{https://github.com/Bala93/Context-aware-segmentation}{https://github.com/Bala93/Context-aware-segmentation}}. Each model is trained for 150 epochs using the Adam optimizer, with a learning rate of 1e-4 in all the reported experiments for consistent comparison. $\lambda$ was set empirically to 1 for both $\mathcal{L}_{context}$ and $L_{Seg-GLGAN}$. 

\begin{table}[]
\centering
\caption{Quantitative Results on PROMISE12 and ACDC Dataset.$(Ours^{*})$}
\label{tab:results}
\scriptsize
\begin{tabular}{|l|l|c|c|c|c|}
\hline
\multirow{2}{*}{} & \multirow{2}{*}{} & \multicolumn{2}{c|}{PROMISE 12} & \multicolumn{2}{c|}{ACDC} \\ \cline{3-6} 
 &  & Dice & HD & Dice & HD \\ \hline
\multirow{6}{*}{U-Net} & $\mathcal{L}_{CE}$ & 0.7964 & 9.11 & 0.822 & 5.69 \\ 
 & $\mathcal{L}_{WCE}$ & 0.8374 & 7.82 & 0.8385 & 5.31 \\ 
 & $\mathcal{L}_{DICE}$ & 0.8166 & 8.25 & - & - \\ 
 & $\mathcal{L}_{FOCAL}$ & 0.8269 & 8.00 & 0.8245 & 5.48 \\ 
 & $\mathcal{L}_{CONTEXT}(StROI)^{*}$ & \textbf{0.861} & \textbf{6.91} & \textbf{0.85} & \textbf{4.55} \\ 
 & $\mathcal{L}_{CONTEXT}(DyROI)^{*}$ & 0.861 & 7.30 & 0.8402 & 5.30 \\ \hline
\multirow{3}{*}{GAN} & GAN & 0.7877 & 8.76 & 0.833 & 4.70 \\ 
 & Seg-GLGAN(StROI)$^{*}$ & 0.8019 & 8.40 & 0.835 & 4.69 \\ 
 & Seg-GLGAN(DyROI)$^{*}$ & \textbf{0.86} & \textbf{7.01} & \textbf{0.856} & \textbf{4.38} \\ \hline
\end{tabular}
\end{table}

\begin{figure}
    \centering
    \includegraphics[width=0.7\linewidth]{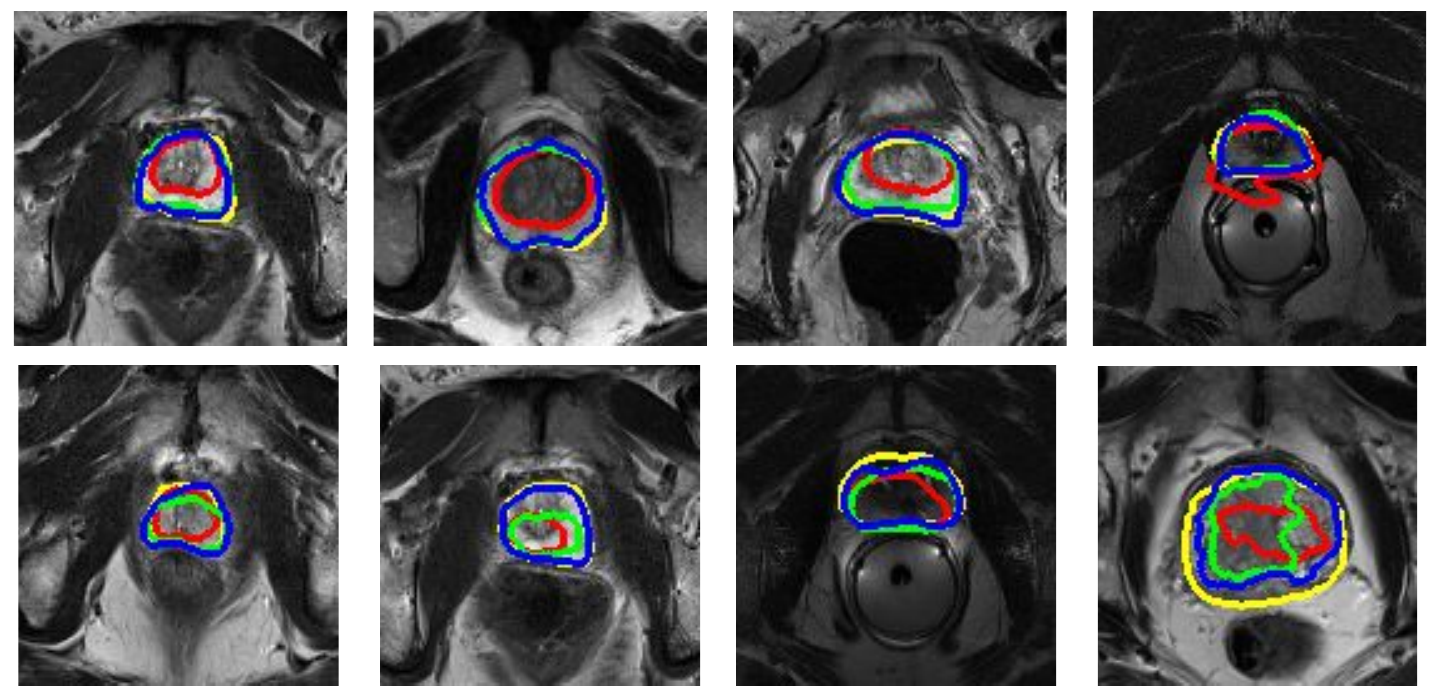}
    \caption{Sample results for binary segmentation(PROMISE12 Dataset), where predictions and targets are overlayed on the inputs. Top Row: Red - U-Net with CE, Green - U-Net with Context CE(StROI), Blue - U-Net with Context CE(DyROI) Yellow - Target. Bottom Row: Red - GAN, Green - Seg-GLGAN(StROI), Blue - Seg-GLGAN(DyROI), Yellow - Target. Best viewed in color.}
    \label{fig:prostate_results}
\end{figure}

\begin{figure}
    \centering
    \includegraphics[width=0.7\linewidth]{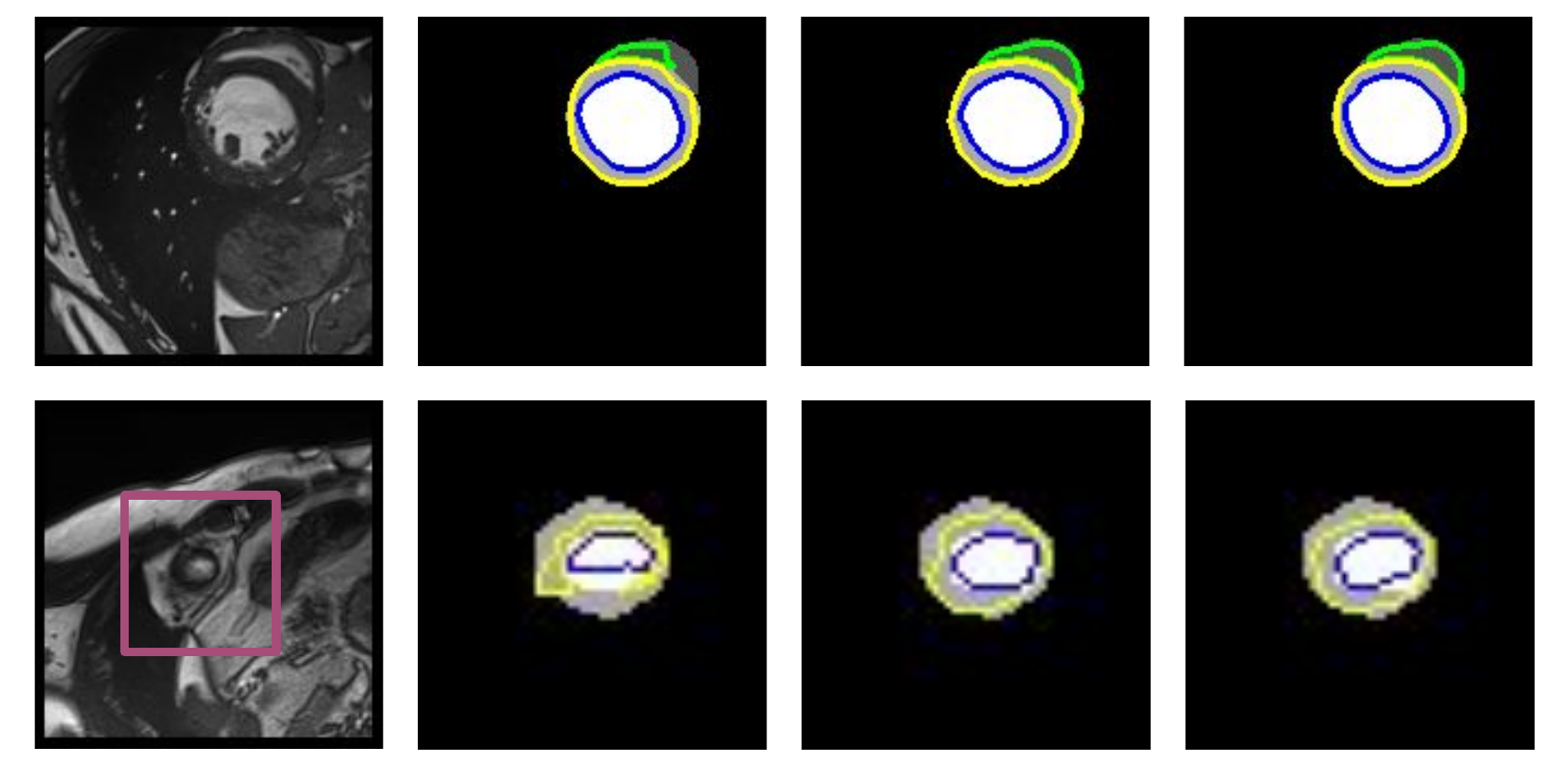}
    \caption{Sample results for multi-class segmentation(ACDC Dataset), where predictions are overlayed on the target. From Left to Right: Top Row - Input, U-Net with CE, U-Net with Context CE(StROI), U-Net with Context CE(DyROI). Bottom Row (The ROI is zoomed for better visualization) - Input, GAN, Seg-GLGAN(StROI), Seg-GLGAN(DyROI). Green(RV), Yellow(Myo) and Blue(LV). Best viewed in color.}
    \label{fig:cardiac_results}
\end{figure}
\subsection{Results and discussion}
The quantitative comparison of proposed context based CE loss ($\mathcal{L}_{CONTEXT}$) with global CE ($\mathcal{L}_{CE}$), weighted CE ($\mathcal{L}_{WCE}$), dice ($\mathcal{L}_{DICE}$) and focal ($\mathcal{L}_{FOCAL}$) for PROMISE12 and ACDC dataset are reported in Table \ref{tab:results}. Similarly, in Table \ref{tab:results}, the quantitative comparison of proposed Seg-GLGAN with GAN is also reported. From the Table, it is evident that our proposed loss $\mathcal{L}_{CONTEXT}$ and network Seg-GLGAN gives better segmentation metrics compared to other popular methods. The significant improvement in performance shown by our methods can be attributed to incorporation of both global and local information to the baseline networks. 
The qualitative comparison is shown in Fig.  \ref{fig:prostate_results} and \ref{fig:cardiac_results}. In Table 1, it can be noted that the Seg-GLGAN (DyROI) performs better than Seg-GLGAN (StROI). This behaviour can be attributed to the usage of dynamic ROI during training. This training procedure might have enabled the Seg-GLGAN (DyROI) network to learn a range of variations in the shape and size of objects. 

The apex and base slices are the challenging images in PROMISE12 dataset as the prostate occupies a relatively smaller region. The baseline methods have difficulty in segmenting prostate in these slices, whereas our approaches have shown promising results. Some cases can be seen in Fig. \ref{fig:prostate_results}. The two major segmentation difficulties in ACDC dataset are: 1) segmentation of structures in apical slices where the structures occupy a relatively smaller region and 2) segmentation of RV as it occupies lesser region compared to other structures. For both these cases, our methods gave better results compared to baseline methods. Sample images displaying the same can be found in Fig. \ref{fig:cardiac_results}. 

The proposed methods can be successfully applied to regular object segmentation tasks because of the ROI. However, the idea of ROI limits the method's extension to sparse segmentation tasks (vessel extraction for retinal images). The future research will be to address this limitation. 





\section{Conclusion}
In this paper, we have proposed a context based cross entropy loss for U-Net and a GAN based network, Seg-GLGAN
to alleviate class imbalance problem in segmentation by considering both global and local context. The proposed methods have shown promising results for challenging datasets.

\newpage

\bibliographystyle{IEEEbib}
\bibliography{isbi}

\end{document}